\documentclass[letterpaper]{article} 
\usepackage{aaai24}  
\usepackage{times}  
\usepackage{helvet}  
\usepackage{courier}  
\usepackage[hyphens]{url}  
\usepackage{graphicx} 
\usepackage{booktabs}
\usepackage{natbib}  
\usepackage{multirow} 
\usepackage{caption} 
\usepackage{algorithm}
\usepackage{algorithmic}
\usepackage{newfloat}
\usepackage{listings}
\DeclareCaptionStyle{ruled}{labelfont=normalfont,labelsep=colon,strut=off} 
\floatstyle{ruled}
\newfloat{listing}{tb}{lst}{}
\floatname{listing}{Listing}
\title{CMUNeXt: An Efficient Medical Image Segmentation Network based on Large Kernel and Skip Fusion}
\author{
    Fenghe Tang\textsuperscript{\rm 1,2}, Jianrui Ding\textsuperscript{\rm 4}, Lingtao Wang\textsuperscript{\rm 4}, Chunping Ning\textsuperscript{\rm 5},  S. Kevin Zhou\textsuperscript{\rm 1,2,3}\thanks{Corresponding author.}\\
}
\affiliations{
    \textsuperscript{\rm 1} School of Biomedical Engineering Division of Life Sciences University of Science and Technology of China (USTC), Hefei, Anhui 230026, China\\
    \textsuperscript{\rm 2} Center for Medical Imaging, Robotics, Analytic Computing and Learning (MIRACLE) Suzhou Institute for Advanced Research University of Science and Technology of China (USTC), Suzhou, Jiangsu 215123, China\\
    \textsuperscript{\rm 3} Key Laboratory of Intelligent Information Processing Institute of Computing Technology Chinese Academy of Sciences (CAS), Beijing, 100190, China\\
    \textsuperscript{\rm 4} School of Computer Science and Technology, Harbin Institute of Technology, Harbin, China\\
    \textsuperscript{\rm 5} Ultrasound Department, The Affiliated Hospital of Qingdao University, Qingdao, China\\
 
}

\usepackage{bibentry}

\begin{document}

\maketitle

\begin{abstract}
The U-shaped architecture has emerged as a crucial paradigm in the design of medical image segmentation networks. However, due to the inherent local limitations of convolution, a fully convolutional segmentation network with U-shaped architecture struggles to effectively extract global context information, which is vital for the precise localization of lesions. While hybrid architectures combining CNNs and Transformers can address these issues, their application in real medical scenarios is limited due to the computational resource constraints imposed by the environment and edge devices. In addition, the convolutional inductive bias in lightweight networks adeptly fits the scarce medical data, which is lacking in the Transformer based network. In order to extract global context information while taking advantage of the inductive bias, we propose CMUNeXt, an efficient fully convolutional lightweight medical image segmentation network, which enables fast and accurate auxiliary diagnosis in real scene scenarios. CMUNeXt leverages large kernel and inverted bottleneck design to thoroughly mix distant spatial and location information, efficiently extracting global context information. We also introduce the Skip-Fusion block, designed to enable smooth skip-connections and ensure ample feature fusion. Experimental results on multiple medical image datasets demonstrate that CMUNeXt outperforms existing heavyweight and lightweight medical image segmentation networks in terms of segmentation performance, while offering a faster inference speed, lighter weights, and a reduced computational cost. The code is available at https://github.com/FengheTan9/CMUNeXt.
\end{abstract}

\section{Introduction}
Medical image segmentation is a crucial but challenging task in computer-aided medical diagnosis, as it can provide doctors with objective and efficient references for lesions and regions of interest. Convolutional Neural Networks (CNNs) have significantly advanced the medical image segmentation tasks, with U-Net \cite{unet} being a representative network that uses an encoder-decoder pyramid architecture for segmentation. U-Net strengthens the transfer ability of knowledge through skip-connections, which can effectively fit scarce medical data. The U-shaped architecture has become a significant paradigm for designing medical image segmentation networks, leading to the proposal of various fully convolutional networks such as Attention U-Net \cite{attunet}, U-Net++ \cite{unet++}, U-Net3+ \cite{unet3+} and CMU-Net \cite{cmunet}. However, the inherent local limitations of ordinary convolution operations make it difficult to extract global information effectively. Recently, hybrid architecture networks based on CNN and Transformer \cite{attention}, such as TransUnet \cite{transunet}, MedT \cite{medT}, and Swin U-Net \cite{swinunet}, have been proposed to combine the inductive bias of CNNs and the global context information extraction ability of Vit \cite{vit}. However, these approaches require significant medical data and computational overhead. While improving the segmentation network's performance is crucial, real medical diagnosis scenarios also require considering factors such as network parameters and compute cost limitations imposed by the environment and edge devices.

The lightweight segmentation network is suitable for deployment on mobile and edge devices, providing doctors and patients with a rapid access to auxiliary diagnosis results. Additionally, edge deployment can address privacy concerns associated with data transmission when using servercentric models. Among these models, UNeXt \cite{unext} has designed a lightweight medical image segmentation network based on a hybrid architecture of CNN and MLP \cite{mlpmixer}, ConvUNeXt \cite{convunext} leverages and improves the ConvNeXt block for efficient segmentation. While most lightweight networks reduce complexity by reducing the width and depth of the network, this approach can have drawbacks, as reducing complexity can result in decreased network performance. Furthermore, global context information is critical for accurately locating lesions in medical images. Although Transformer-based segmentation networks reduce parameters to match mobile device resource constraints, their performance is significantly inferior to that of lightweight CNNs. The inductive bias of convolution can better fit scarce medical data, which is lacking in the ViT network.

Recently, several proposed works have utilized CNN with large convolution kernels as an alternative to Transformer for extracting global context information \cite{convmixer, 31x31, convnext}. For instance, ConvMixer \cite{convmixer} uses large convolution kernels to mix distant spatial position information, while RepLKNet \cite{31x31} extracts global information using a 31x31 super large convolution kernel. ConvNeXt \cite{convnext} use large convolution kernels and Swin Transformer \cite{swin} architecture design tricks to improve the performance of fully convolutional networks in all aspects. Convolutions with a large kernel can take advantage of convolution inductive bias while gaining larger receptive field information. However, the above work is only utilized for general natural images, the huge gap between natural images and medical images makes it difficult to apply in medical scenarios. To respond the above issue and make it meet the real diagnostic application needs, we propose CMUNeXt, a fully convolutional lightweight medical image segmentation network that follows the U-shaped architecture design, consisting of a five-level encoder-decoder structure with skip-connections. We redesign each module of the network to achieve optimal performance while keeping the weight lighter. 

To effectively extract global context information while reducing redundant parameters, inspired by the ConvMixer \cite{convmixer} and ConvNeXt \cite{convnext}, we propose the CMUNeXt block in the encoder stage. The CMUNeXt block replaces ordinary convolutions with depthwise convolutions with large kernels and two pointwise convolutions with inverted bottleneck designs to fully mix distant spatial and location information. In the decoder stage with skip-connections, the Skip-Fusion block is proposed to achieve smooth skip-connections. It uses grouped convolution with two inverted bottleneck designed pointwise convolutions to replace ordinary convolutions for sufficient fusion between encoder and decoder semantic features. We select four ultrasound medical image datasets of breast and thyroid to evaluate CMUNeXt. Extensive experimental results demonstrate that CMUNeXt achieves a better trade-off between segmentation performance and computational consumption than other advanced and widely used segmentation methods. It gains better segmentation performance with fewer parameters, lower computational consumption and quicker inference time.

It is worth noting that the reason we selected ultrasound datasets for evaluation is the position of the lesions is not fixed and characteristics of the lesions exhibit multi-scale and large morphological differences, which can better illustrate the effectiveness of our proposed method in extracting global context information of medical images. This work contributes in the following ways: 
\begin{itemize}
\item We propose CMUNeXt, a lightweight fully convolutional medical image segmentation network that effectively extracts global context information in medical images and achieves fast and precise segmentation.
\item We introduce the CMUNeXt block for extracting global information, which can fully extract and mix distant spatial and position information in medical images with minimal parameters.
\item We propose the Skip-Fusion block in the skip-connections for smooth feature fusion, which enables efficient transfer of encoder knowledge to the decoder.
\item We conduct extensive experiments to demonstrate that CMUNeXt achieves a better trade-off between the best segmentation performance and computational consumption. 
\end{itemize}

Additionally, we have open-sourced the CMUNeXt and released a U-shaped architecture-based medical image segmentation benchmarks at: \url{https://github.com/FengheTan9/Medical-Image-Segmentation-Benchmarks}. 

\section{Method}
\begin{figure*}[ht]
\centering
\includegraphics[width=0.9\textwidth]{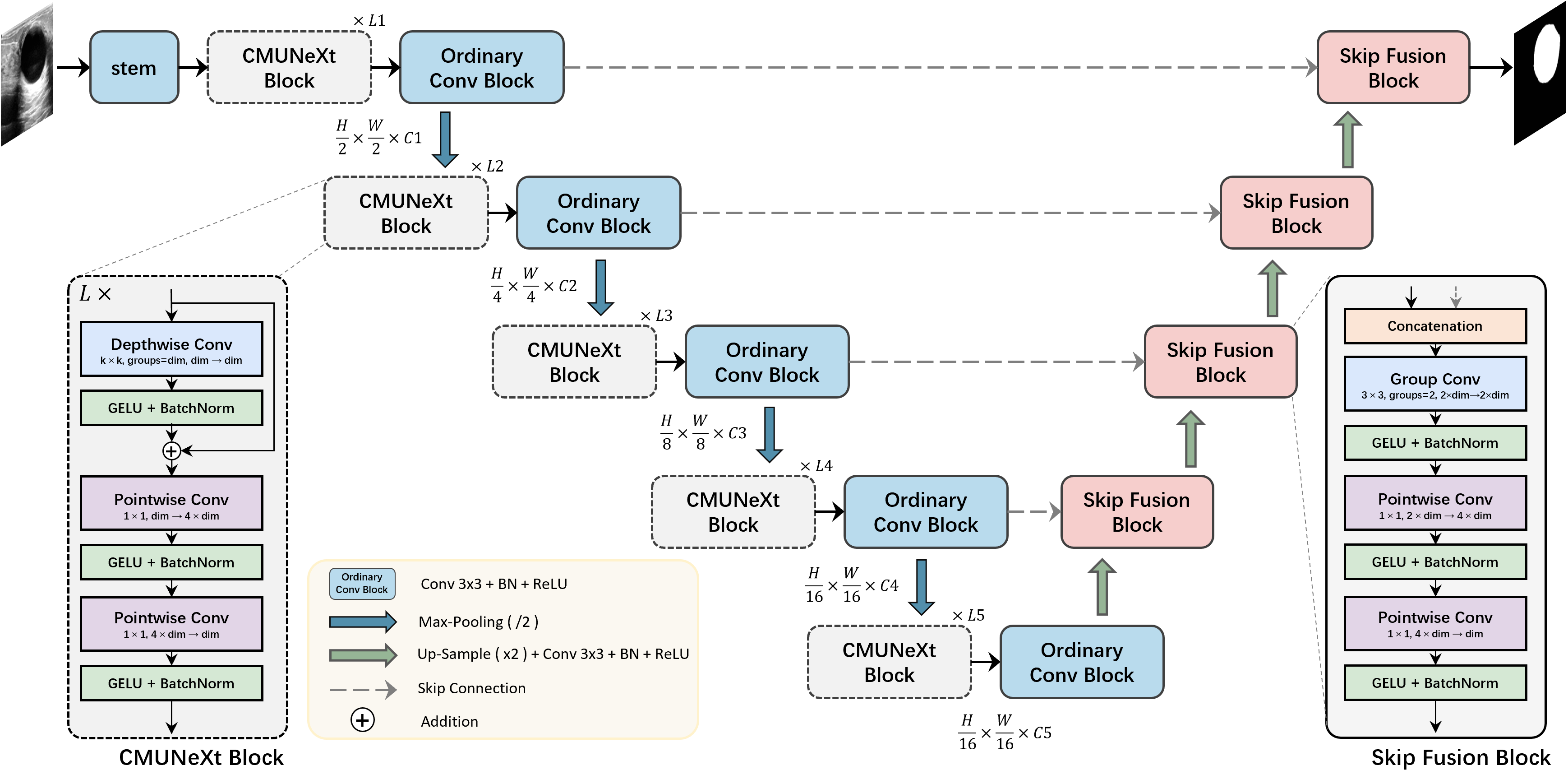} 
\caption{Overview of CMUNeXt.}
\label{fig1}
\end{figure*}
\subsection{Architecture Overview}
The architecture of our proposed CMUNeXt is illustrated in Figure 1. It consists of five layers from top to bottom and it is divided into two stages: the encoder stage and the decoder stage with skip-connections. In the encoder stage, the CMUNeXt block is employed to extract different levels of global context information, followed by an ordinary convolution block to expand the number of channels. In the decoder stage with skip-connections, the Skip-Fusion block is used to fully fuse global semantic features from the encoder with up-sampled features from the decoder. The length of the CMUNeXt block in each layer is denoted as L1 to L5 from top to bottom, and the kernel size of each block is K1 to K5, with the number of channels C1 to C5.

\subsection{Encoder Stage}
As depicted in Figure 1, the encoder comprises five levels from top to bottom, each of which consists of a CMUNeXt block, an ordinary convolution block and a down-sampling operation. In addition, we utilize stem to extract the original features from the input image on the top level.
\subsubsection{Stem.} The stems design in networks such as ResNet \cite{resnet} and ConvNeXt \cite{convnext} can reduce resolution of outputs and cause inconsistency with the skip-connection of the top level. To avoid this, we adopt an ordinary convolution block, which is equipped with a convolution layer, a batch normalization layer, and a ReLU activation, with a kernel size of 3$\times$3, a stride of 1, and a padding of 1. By using this approach, we can maintain consistency with the skip-connection of the top level and ensure that important features are not lost during the encoding process.
\subsubsection{CMUNeXt Block.} The core component of CMUNeXt Block is depthwise separable convolution, which has been popularized by MobileNetv2 \cite{mobilenetv2} and MobileVit \cite{mobilevit}. Depth separable convolution uses a combination of depthwise convolution (i.e., the number of groups is equal to the channels) and pointwise convolution (i.e., kernel size 1$\times$1) to replace a complete ordinary convolution operation. The separated depthwise convolution is used to extract the spatial dimension information, and the subsequent pointwise convolution leads to a separation of spatial and channel mixing. Compared with ordinary convolution, depthwise convolution can effectively reduce network parameters and computational cost. In the CMUNeXt Block, we use a depthwise convolution of a large kernel size to extract the global information of each channel, followed by a residual connection. To fully mix spatial and channel information, we apply two pointwise convolutions after the depthwise convolution and perform an inverted bottleneck design on them. This design involves setting the hidden dimension between the two pointwise convolutional layers four times wider than the input dimension. The inverted bottleneck design has been further generalized in ConvNeXt \cite{convnext}. The expanded hidden dimension can comprehensively and fully mix the global spatial dimension information extracted by the depthwise convolution. Furthermore, we utilize GELU activation and post-activation BatchNorm layer after each of the convolutions. The CMUNeXt Block is defined as:
\begin{equation}
f_l^\prime=BN\left(\sigma_1\left\{DepthwiseConv\left(f_{l-1}\right)\right\}\right)+f_{l-1}
\end{equation}
\begin{equation}
f_l^{\prime\prime}=BN\left(\sigma_1\left\{PointwiseConv\left(f_{l}^\prime\right)\right\}\right)
\end{equation}
\begin{equation}
f_l=BN\left(\sigma_1\left\{PointwiseConv\left(f_l^{\prime\prime}\right)\right\}\right)
\end{equation}
where $f_l$ represents the output feature map of layer l in the ConvMixer block, $\sigma_1$ represents the GELU activation, and BN represents Batch Normalization layer. Since the input and output feature maps of CMUNeXt Block maintain the same resolution and channel size, we use an ordinary convolutional block to expand the channel size by two times.
\subsubsection{Down Sample.} Many methods, such as ConvNeXt \cite{convnext} and MobileVit \cite{mobilevit}, utilize convolution with a stride of 2 for spatial downsampling on natural images. Introducing a standalone downsampling layer can enhance the stability during training and simultaneously boost performance. However, medical images often exhibit low resolution and minor local edge variations. In comparison to using convolution for down sampling, the conventional pooling operation can efficiently filter out the noise present in medical images while maintaining a minimum computational overhead. Consequently, the down sampling strategy adopted in CMUNeXt is max pooling, employing a filter window of 2$\times$2 and a stride of 2.

\subsection{Decoder Stage with skip-connections}
The decoder is also composed of five levels arranged from bottom to top. Each level comprises a Skip-Fusion block and an upsampling block.
\subsubsection{Skip-Fusion Block.} Traditional skip-connections typically use ordinary convolution operations for feature fusion, which can be direct and blunt, and it increases the burden on both the encoder and decoder. In our Skip-Fusion block, we utilize group convolution as the core component to address these issues. We divide the convolution operation into two groups, and perform "feature by feature" extraction on the skipped encoder feature and up-sampled decoder feature, respectively. The kernel size of group convolution is 3$\times$3, with a stride of 1 and padding of 1. To enable sufficient feature fusion, we incorporate two inverted bottleneck pointwise convolutions after the group convolution. Skip-Fusion block assigns the feature adaptation before fusion to the group convolution.  And the efficient and dense pointwise convolution undertakes heavy feature fusion work. Additionally, each of the convolutions in the Skip-Fusion block is followed by a GELU activation and BatchNorm layer. The definition of Skip-Fusion Block is as follows:
\begin{equation}
f_{concat}=Concat{
\left(
\begin{array}{l}
BN{\{}OridinaryConv(f_{\mathcal{E}}){\}}, \\
BN{\{}OridinaryConv(f_{\mathcal{D}}){\}}
\end{array} 
\right)
}
\end{equation}
\begin{equation}
f_{fusion}^{\prime}=BN\left(\sigma_1\left\{PointwiseConv\left(f_{concat}\right)\right\}\right)
\end{equation}
\begin{equation}
f_{fusion}=BN\left(\sigma_1\left\{PointwiseConv\left(f_{fusion}^{\prime}\right)\right\}\right)
\end{equation}
where $f_{fusion}$ represents the ouput fusion feature map in Skip-Fusion block, $f_\mathcal{E}$  and $f_\mathcal{D}$ represents the encoder and decoder features, respectively.
\subsubsection{Upsampling Block.} The upsampling Block consists of an upsampling layer, a convolutional layer, a batch normalization layer, and a ReLU activation function. We utilize a bilinear interpolation to upsample the feature maps by a factor of two. The convolution layer has a kernels size of 3×3, with a stride of 1 and padding of 1. By employing these techniques, we are able to effectively increase the resolution of the feature maps while preserving important features.

\section{Experiments}
\subsection{Datasets} 
\subsubsection{The BUS dataset.} The Breast UltraSound (BUS) \cite{busis} public dataset contains 562 breast ultrasound images collected using five different ultrasound devices, including 306 benign cases and 256 malignant cases, each with corresponding ground truth labels.

\subsubsection{The BUSI dataset.} The Breast UltraSound Images (BUSI) \cite{busi} public dataset collected from 600 female patients, includes 780 breast ultrasound images, covering 133 normal cases, 487 benign cases, and 210 malignant cases, each with corresponding ground truth labels. Following recent studies \cite{unext, global}, we only utilize benign and malignant cases from this dataset.
\subsubsection{The TNSCUI dataset.} The Thyroid Nodule Segmentation and Classification in Ultrasound Images 2020 (TNSCUI) (\url{https://tn-scui2020.grand-challenge.org/Home/}) public dataset was collected by the Chinese Artificial Intelligence Alliance for Thyroid and Breast Ultrasound (CAAU). It includes 3644 cases of different ages and genders, each with corresponding ground truth labels.
\subsubsection{The TUS dataset.} The Thyroid UltraSound (TUS) private dataset was collected using three different ultrasound machines from the Ultrasound Department of the Affiliated Hospital of Qingdao University. It includes 192 cases, with a total of 1942 thyroid ultrasound images and corresponding segmentation ground truth labels by three experienced radiologists.

We randomly split each of the four medical image datasets into training and validation sets three times, using a 70/30 split for training and validation, respectively. This allows us to assess the generalization performance of our proposed method across different datasets and random splits.

\subsection{Implementation Details}
\subsubsection{CMUNeXt Variants Setting.} To evaluate the effectiveness of our proposed CMUNeXt architecture, we construct different variants of the model with varying numbers of channels (C), block lengths (L) per level, and kernel sizes (K). Specifically, we construct CMUNeXt-S and CMUNeXt-L variants, with compute ratios of (1,1,1,1) and (1,1,6,3), respectively. Additionally, we modify the kernel size of the bottom block of CMUNeXt-S to 9, in order to compensate for the smaller global receptive field resulting from the reduction in compute ratio. By constructing these different variants, we can evaluate the impact of varying model parameters on the performance of the CMUNeXt architecture, and determine the optimal deployment configuration for different real-world medical scenarios.

\begin{table}[t]
\renewcommand{\arraystretch}{1.2} 

\resizebox{\linewidth}{!}{
\centering
\begin{tabular}{c|ccccc|ccccc|ccccc}

\toprule[1.1pt]

\multirow{2}{*}[-0.5em]{Networks} & \multicolumn{5}{c|}{number of channels} & \multicolumn{5}{c|}{length of blocks} & \multicolumn{5}{c}{kernel size}\\
\cmidrule{2-16}

\multicolumn{1}{c|}{}  & C1 &  C2 & C3 & C4 & C5 & L1 &  L2 & L3 & L4 & L5 & K1 &  K2 & K3 & K4 & K5 \\

\midrule[0.7pt]

\multicolumn{1}{c|}{CMUNeXt-S}  & 8 &  16 & 32 & 64 & 128 & 1 &  1 & 1 & 1 & 1 & 3 &  3 & 7 & 7 & 9 \\

\midrule[0.7pt]

\multicolumn{1}{c|}{CMUNeXt}  & 16 &  32 & 128 & 160 & 256 & 1 &  1 & 1 & 3 & 1 & 3 &  3 & 7 & 7 & 7 \\

\midrule[0.7pt]

\multicolumn{1}{c|}{CMUNeXt-L}  & 32 &  64 & 128 & 256 & 512 & 1 &  1 & 1 & 6 & 3 & 3 &  3 & 7 & 7 & 7 \\

\bottomrule[1.1pt]
\end{tabular}
}
\caption{CMUNeXt variants.}
\label{table1}
\end{table}

\subsubsection{Training and Hyperparameter Setting.} We implement CMUNeXt by the PyTorch. The loss $\mathcal{L}$ between the predicted ${\hat{y}}$ and ground truth y is defined as a combination of binary cross entropy (BCE) and dice loss (Dice):
\begin{equation}
\mathcal{L}=0.5\times BCE\left(\hat{y},y\right)+Dice\left(\hat{y},y\right)
\label{eq}\end{equation}

We resize all training cases of four datasets to 256$\times$256 and apply random rotation and flip for simple data augmentations. In addition, we use the SGD optimizer with a weight decay of 1e-4 and a momentum of 0.9 to train the networks. The initial learning rate is set to 0.01, and the poly strategy is used to adjust the learning rate. The batch size is set to 8 and the training epochs are 300. All the experiments are conducted using a single NVIDIA GeForce RTX4090 GPU.

\subsection{Evaluation Metrics and Comparison Methods} We adopt Intersection over Union (IoU) and F1-score as well as number of parameters (in M), computational complexity (in GFLOPs) and Frames Per Second (in FPS) to evaluate the performance of different segmentation models comprehensively. We compare the performance of CMUNeXt with nine widely used and state-of-the-art medical image segmentation networks, including convolutional based networks such as U-Net \cite{unet}, Attention U-Net \cite{attunet}, U-Net++ \cite{unet++}, U-Net3+ \cite{unet3+}, CMU-Net \cite{cmunet}, hybrid CNN and Transformer based networks such as TransUnet \cite{transunet}, MedT \cite{medT}, Swin-Unet \cite{swinunet}, and lightweight networks such as UNeXt \cite{unext} and ConvUNeXt \cite{convunext}.
\begin{table*}[htb]
\renewcommand{\arraystretch}{1.2} 

\resizebox{\linewidth}{!}{
\centering
\begin{tabular}{cccc|cc|cc|cc|cc}

\toprule[1.1pt]

\multirow{3}{*}[-0.5em]{Networks}&\multirow{3}{*}[-0.5em]{Params\ (M)$\downarrow$}&\multirow{3}{*}[-0.5em]{GFLOPs$\downarrow$}&\multirow{3}{*}[-0.5em]{FPS$\uparrow\ \ $} & \multicolumn{8}{c}{Metrics\ (\%)}\\
\cmidrule{5-12}

\multicolumn{4}{c|}{} & \multicolumn{2}{c|}{BUS}  & \multicolumn{2}{c|}{BUSI}   & \multicolumn{2}{c|}{TUS} & \multicolumn{2}{c}{TUSCNI}         \\ 
\cmidrule{5-12}
 
\multicolumn{4}{c|}{}   & \ IoU$\uparrow$ & \multicolumn{1}{c|}{\ F1$\uparrow$} & \ IoU$\uparrow$  & \multicolumn{1}{c|}{\ F1$\uparrow$} & \ IoU$\uparrow$ & \multicolumn{1}{c|}{\ F1$\uparrow$} & \ IoU$\uparrow$ & \multicolumn{1}{c}{\ F1$\uparrow$}    \\ 
\midrule[0.7pt]

\multicolumn{1}{c}{U-Net} & 34.52 & 65.52 & 139.32$\ \ $ & 86.73 & 92.46 & 68.61 & 76.97 & 82.60 & 89.82 & 75.73 & 84.11\\ 

\multicolumn{1}{c}{Attention U-Net} & 34.87 & 66.63 & 129.92$\ \ $ & 86.79 & 92.55 & 68.55 & 76.88 & 82.68 & 89.86 & 75.83 & 84.28\\ 

\multicolumn{1}{c}{U-Net++} & 26.90 & 37.62 & 125.50$\ \ $ & 86.92 & 92.74 & 69.49 & 78.06 & 82.54 & 89.88 & 76.90 & 85.13\\ 

\multicolumn{1}{c}{U-Net3+}  &  26.97 & 199.74 & 50.60$\ \ $ & 86.48 & 92.34 & 68.38 & 76.88 & 82.34 & 89.65 & 75.52 & 83.93\\ 

\multicolumn{1}{c}{CMU-Net}  &  49.93 & 91.25 & 93.19$\ \ $ & \underline{87.18} & \underline{92.89} & \underline{71.42} & 79.49 & \underline{83.10} & \underline{90.16} & 77.12 & 85.35\\ 

\midrule[0.7pt]

\multicolumn{1}{c}{TransUnet}  & 105.32 & 38.52 & 112.95$\ \ $ & 86.73 & 92.44 & 71.39 & \underline{79.85} & 82.19 & 89.48 & \underline{77.63} & \underline{85.76}\\ 

\multicolumn{1}{c}{MedT} & 1.37 & 2.40 & 22.97$\ \ $ & 80.81 & 88.78 & 63.36 & 73.37 & 80.45 & 88.56 & 71.00 & 80.87\\ 

\multicolumn{1}{c}{SwinUnet} & 27.14 & 5.91 & 392.21$\ \ $ & 85.27 & 91.99 & 63.59 & 76.94 & 81.03 & 89.41 & 75.77 & 85.82\\ 

\midrule[0.7pt]

\multicolumn{1}{c}{UNeXt} & \underline{1.47} & \underline{\textbf{0.58}} & \underline{650.48}$\ \ $ & 84.73 & 91.20 & 65.04 & 74.16 & 79.99 & 87.99 & 71.04 & 80.46\\ 

\multicolumn{1}{c}{ConvUNeXt} & 3.50 & 7.25 & 457.77$\ \ $ & 85.65 & 91.89 & 67.05 & 75.96 & 81.66 & 89.27 & 75.45 & 85.82\\ 

\midrule[0.7pt]

\multicolumn{1}{c}{CMUNeXt-S} & \textbf{0.41} & 1.09 & \textbf{657.01}$\ \ $ & 85.82 & 91.93 & 68.84 & 77.33 & 81.69 & 89.20 & 74.20 & 82.99\\ 

\multicolumn{1}{c}{CMUNeXt} & 3.14 & 7.41 & 471.43$\ \ $ & 87.02 & 92.72 & \textbf{71.56} & \textbf{79.86} & 82.91 & 90.15 & 77.01 & 85.21\\ 

\multicolumn{1}{c}{CMUNeXt-L} & 8.28 & 17.18 & 281.87$\ \ $ & \textbf{87.44} & \textbf{92.98} & 71.18 & 79.66 & \textbf{83.13} & \textbf{90.29} & \textbf{77.83} & \textbf{85.84}\\ 

\midrule[0.7pt]
\multicolumn{4}{c|}{Improvement} & +0.26 & +0.09 & +0.14 & +0.01 & +0.03 & +0.13 & +0.20 & +0.08\\

\bottomrule[1.1pt]
\end{tabular}
}
\caption{Comparison Result on Medical Ultrasound Datasets.}
\label{table2}
\end{table*}

\begin{figure*}[t]

\begin{minipage}[b]{0.33\linewidth}
  \centering
  \centerline{\includegraphics[width=\columnwidth]{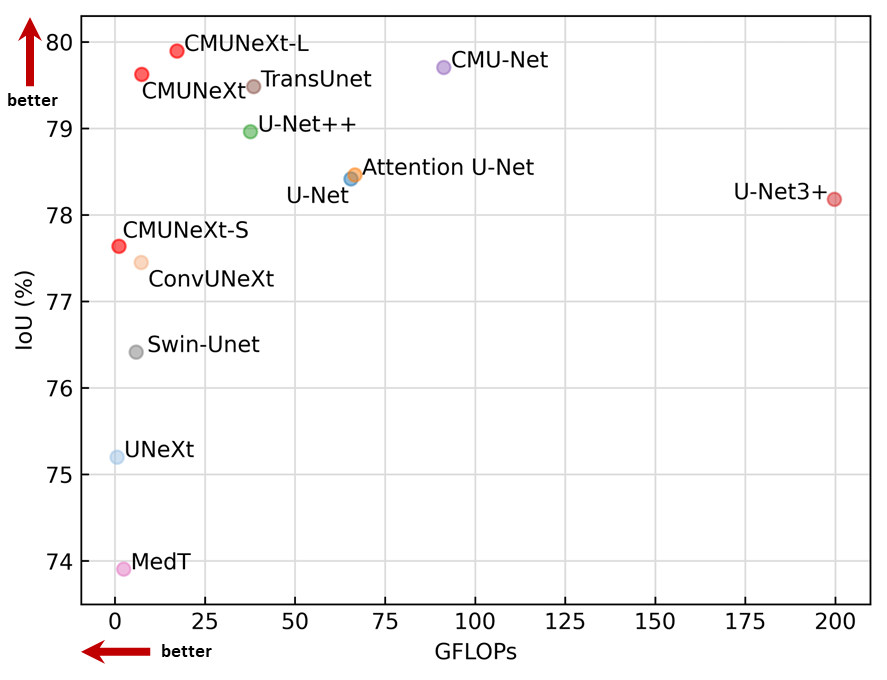}}
\end{minipage}
\begin{minipage}[b]{0.33\linewidth}
  \centering
  \centerline{\includegraphics[width=\columnwidth]{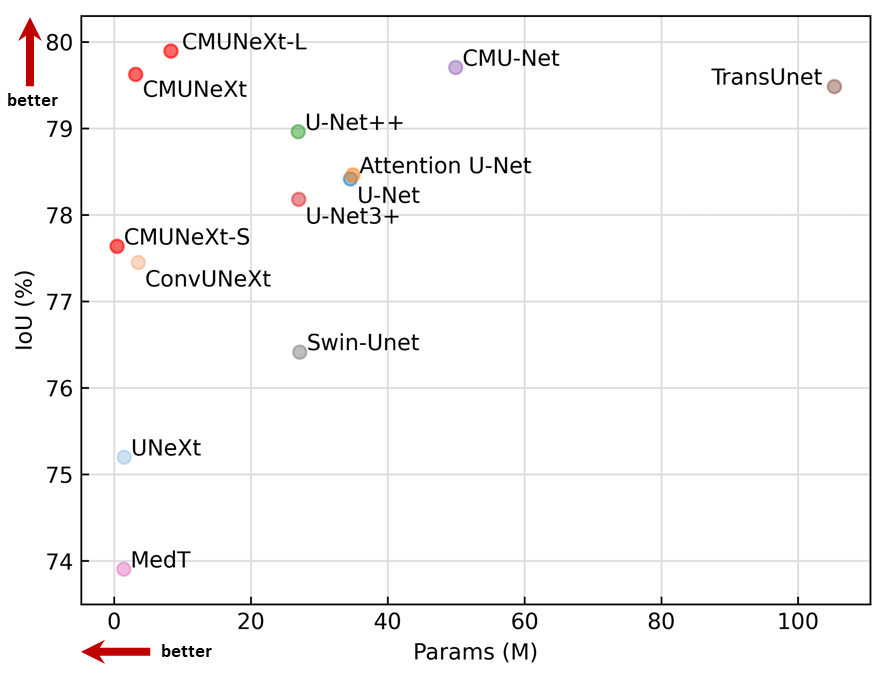}}
\end{minipage}
\begin{minipage}[b]{0.33\linewidth}
  \centering
  \centerline{\includegraphics[width=\columnwidth]{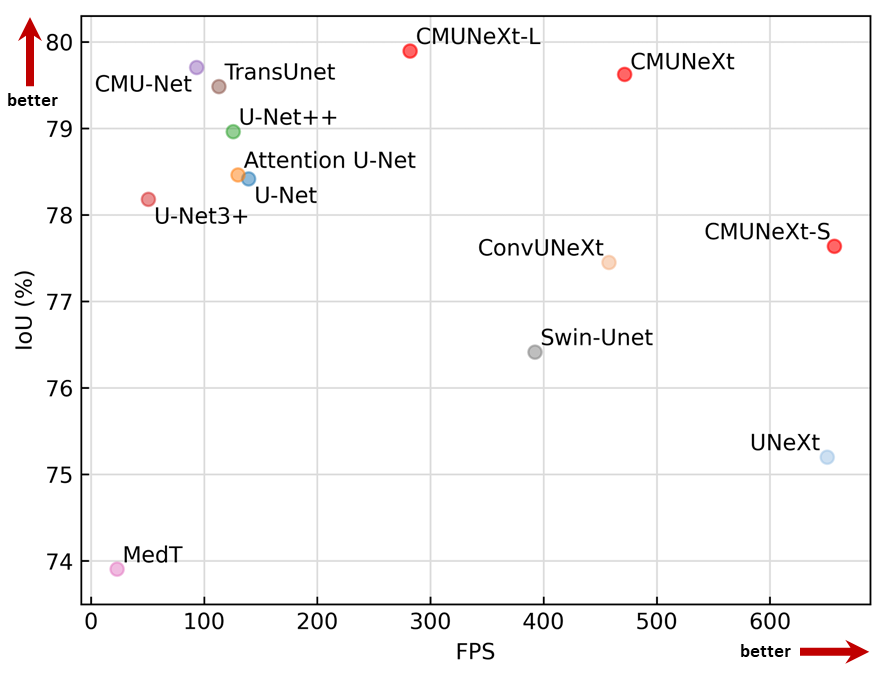}}
\end{minipage}

\caption{Comparison Charts.}
\label{fig2}
\end{figure*}

\begin{table*}[t]
\renewcommand{\arraystretch}{1.2} 

\resizebox{\linewidth}{!}{
\centering
\begin{tabular}{cccc|cc|cc|cc|cc}

\toprule[1.1pt]

\multirow{3}{*}[-0.5em]{Networks}&\multirow{3}{*}[-0.5em]{Params\ (M)$\downarrow$}&\multirow{3}{*}[-0.5em]{GFLOPs$\downarrow$}&\multirow{3}{*}[-0.5em]{FPS$\uparrow\ \ $} & \multicolumn{8}{c}{Metrics\ (\%)}\\
\cmidrule{5-12}

\multicolumn{4}{c|}{} & \multicolumn{2}{c|}{BUS}  & \multicolumn{2}{c|}{BUSI}   & \multicolumn{2}{c|}{TUS} & \multicolumn{2}{c}{TUSCNI}         \\ 
\cmidrule{5-12}
 
\multicolumn{4}{c|}{}   & \ IoU$\uparrow$ & \multicolumn{1}{c|}{\ F1$\uparrow$} & \ IoU$\uparrow$  & \multicolumn{1}{c|}{\ F1$\uparrow$} & \ IoU$\uparrow$ & \multicolumn{1}{c|}{\ F1$\uparrow$} & \ IoU$\uparrow$ & \multicolumn{1}{c}{\ F1$\uparrow$}    \\ 
\midrule[0.7pt]

\multicolumn{1}{c}{Original U-Net} & 34.52 & 65.52 & 139.32$\ \ $  & 86.73 & 92.46 & 68.61 & 76.97 & 82.60 & 89.82 & 75.73 & 84.11\\ 

\multicolumn{1}{c}{Reduced U-Net} & 3.34 & 7.34 & 536.58$\ \ $  & 86.55 & 92.37 & 67.97 & 76.35 & 82.21 & 89.52 & 75.30 & 83.80\\ 

\multicolumn{1}{c}{CMUNeXt Block} & 3.18 & 6.81 & 499.52$\ \ $ & 87.13 & 92.84 &	70.35 & 78.85 & 82.65 & 90.01 & 76.45 & 84.82\\ 

\multicolumn{1}{c}{CMUNeXt Block + Skip-Fusion Block} & 3.14 & 7.41 & 471.43$\ \ $ & 87.02 & 92.72 & 71.56 & 79.86 & 82.91 & 90.15 & 77.01 & 85.21\\ 

\bottomrule[1.1pt]
\end{tabular}
}
\caption{Ablation Study on CMUNeXt.}
\label{table3}
\end{table*}

\begin{figure*}[t]
\centering
\includegraphics[width=0.98\textwidth]{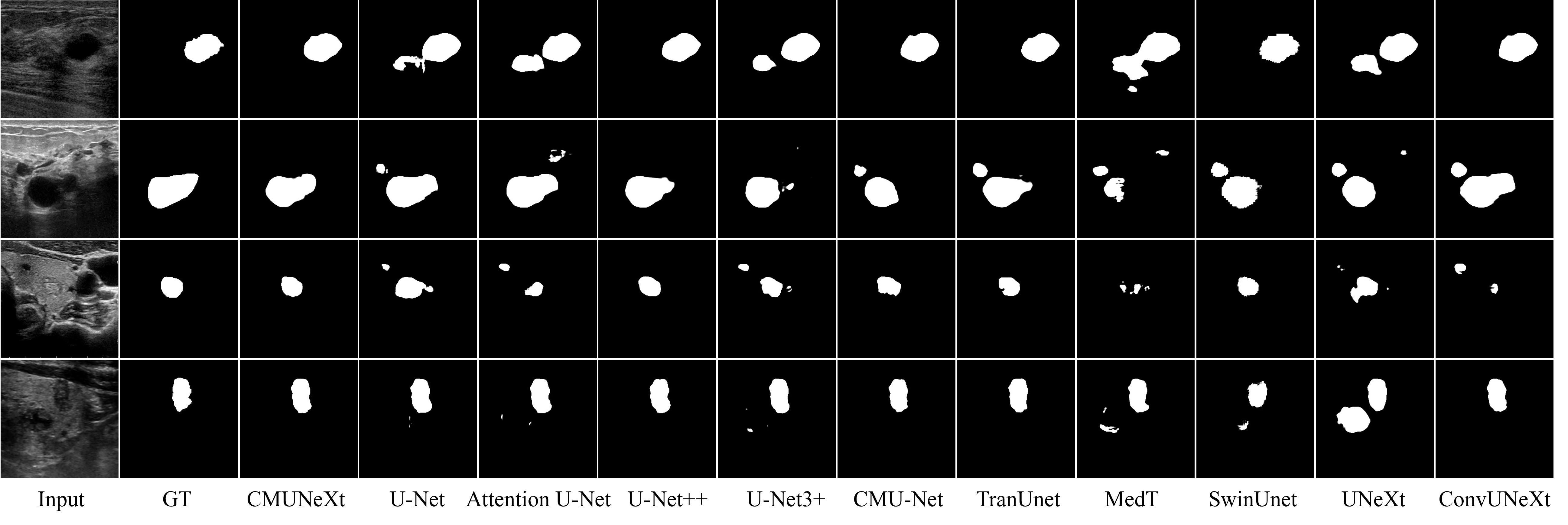} 
\caption{Visualization Results. Row 1 - BUS sample, Row 2 - BUSI sample, Row 3 – TUS sample, Row 4 TNSCUI sample.}
\label{fig3}
\end{figure*}

\section{Discussion}

\subsection{Experiment Results} The comparison results of our proposed CMUNeXt with recent and widely used networks are summarized in Table 2. Overall, our proposed CMUNeXt architecture achieves state-of-the-art segmentation performance while maintaining a reasonable computational complexity. In particular, on the BUSI dataset, CMUNeXt achieves the best performance, an IoU of 71.56 and F1 scores of 79.86.

In addition, our proposed CMUNeXt-L achieves state-of-the-art results on BUS, TUS and TNSCUI datasets. Compared with advanced heavy-weight networks such as the CMU-Net and TransUNet, CMUNeXt-L achieves a reduction in parameters and GFLOPs by 6.03x and 2.24x, respectively, while increasing FPS by at least 2.49x. It is proved that our proposed method effectively improves the global representation ability of the network while reducing redundant parameters.

It is worth noting that even our lightest CMUNeXt-S achieves competitive segmentation performance. Compared with the lightweight segmentation network UNeXt, CMUNeXt-S achieves a respective increase of 1.1-3.8\% in IoU and 0.7-3.2\% in F1 score on the four datasets, while obtains fewer parameters (0.41M) and achieving a higher inference frame (657 FPS). This highlights the importance and effectiveness of using a large kernel to compensate for the local limitations of the convolution's inductive bias, and the potential of convolution-based models to achieve high performance gains.

Furthermore, we plot a comparison chart of IoU with FPS, GFLOPs, and parameters by averaging the segmentation performance on the four datasets. As shown in Figure 2, CMUNeXt achieves the best segmentation performance while achieving faster inference speed, lighter weight, and lower GFLOPs compared to other models. These results demonstrate that CMUNeXt meets the performance and environmental requirements of edge and mobile device deployment and operation. In real application scenarios, different variants of CMUNeXt can be selected based on specific needs, to achieve the best trade-off between accuracy and inference cost.

\subsection{Ablation Study}
To thoroughly evaluate the effectiveness of our proposed CMUNeXt, we conduct comprehensive ablation experiments on the four datasets to analyze the contribution of each block. The evaluation results are presented in Table 3.

First, we reduce the number of channels of the original U-Net (C1=64, C2=128, C3=256, C4=512, C5=1024) to obtain the Reduced U-Net (C1=16, C2=32, C3 =128, C4=160, C5=256). Surprisingly, we find that the segmentation performance only drops slightly even though the number of parameters and computation in the network is drastically reduced. This suggests that the original U-Net architecture has a lot of parameter redundancy, and increasing the network width cannot substantially help improve the performance, and increasing the network width may not significantly improve the performance. 

Next, we introduce the CMUNeXt block into the Reduced U-Net architecture. We observe that the segmentation performance is greatly improved, while the parameters and GFLOPs is further reduced, and the FPS is further improved. This is because the CMUNeXt block uses a large convolution kernel to extract global context information, while reducing redundant network parameters, resulting in a more efficient and effective feature representation. 

Finally, we introduce the Skip-Fusion block on this basis. We observe that the segmentation performance is further improved, which demonstrates that the Skip-Fusion module can effectively enhance the knowledge transfer capability of the network.

Overall, our ablation experiments demonstrate that each block in the proposed CMUNeXt architecture contributes significantly to the segmentation performance improvements. The results also highlight the importance of using large convolution kernels to extract global context information and reducing redundant network parameters to improve the efficiency and effectiveness of the network.

\subsection{Visualization of Results}
We validate the effectiveness of our proposed method by visualizing segmentation results with state-of-the-art methods in Figure 3. Our method achieves more accurate spatial localization and lesion shape.  For the case with multiple microcalcifications (Row 3) and low contrast (Row 4), CMUNeXt achieves more accurate lesion area and shape by learning the global positional context relationship of lesions, organs and tissues.

\subsection{Training Stability Assessment}
To assess the training stability of CMUNeXt, we compare it with the benchmark U-Net (blue) and the lightweight network UNeXt (purple). We train these models from scratch on different medical image datasets, and the training curves are shown in Figure 4. It demonstrates that CMUNeXt (orange) achieves more stable training results compared to U-Net and UNeXt. Furthermore, CMUNeXt shows faster convergence speed, which is particularly evident in the early stages of training.

\begin{figure}[t]
\begin{minipage}[b]{0.495\linewidth}
  \centering
  \centerline{\includegraphics[width=\columnwidth]{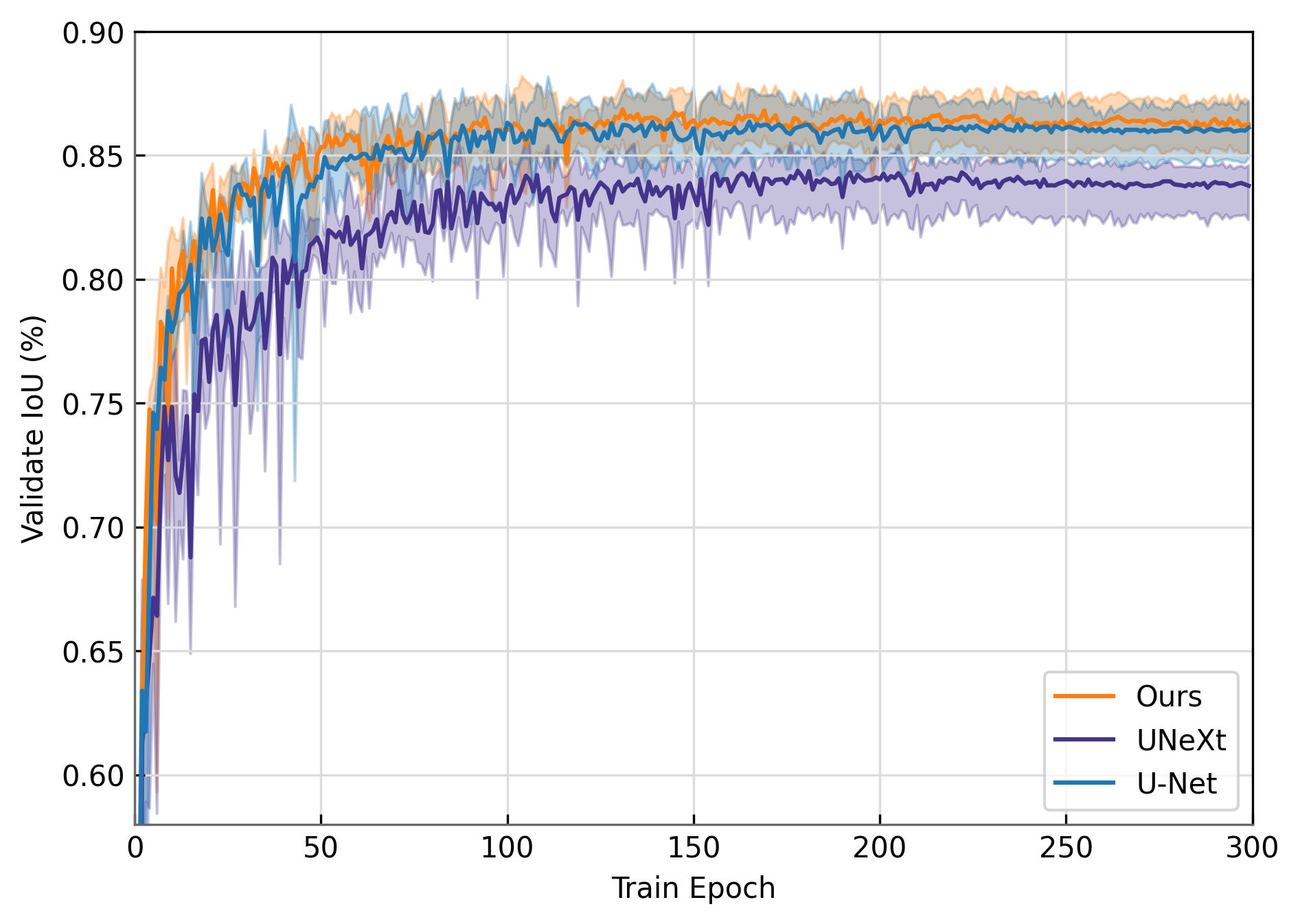}}
  \centerline{\scriptsize{(a) BUS}}\medskip
\end{minipage}
\begin{minipage}[b]{0.495\linewidth}
  \centering
  \centerline{\includegraphics[width=\columnwidth]{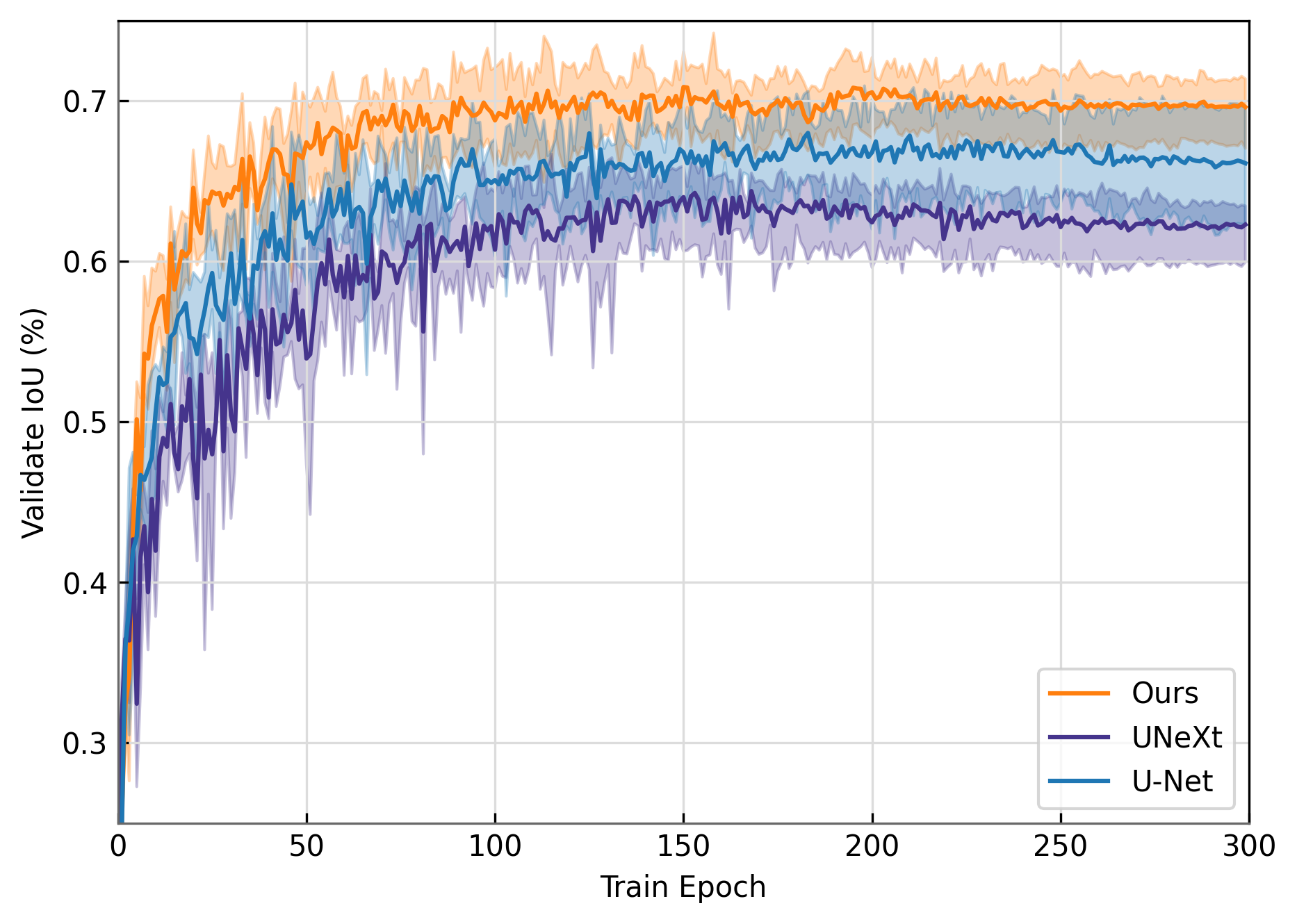}}
  \centerline{\scriptsize{(b) BUSI}}\medskip
\end{minipage}
\begin{minipage}[b]{0.495\linewidth}
  \centering
  \centerline{\includegraphics[width=\columnwidth]{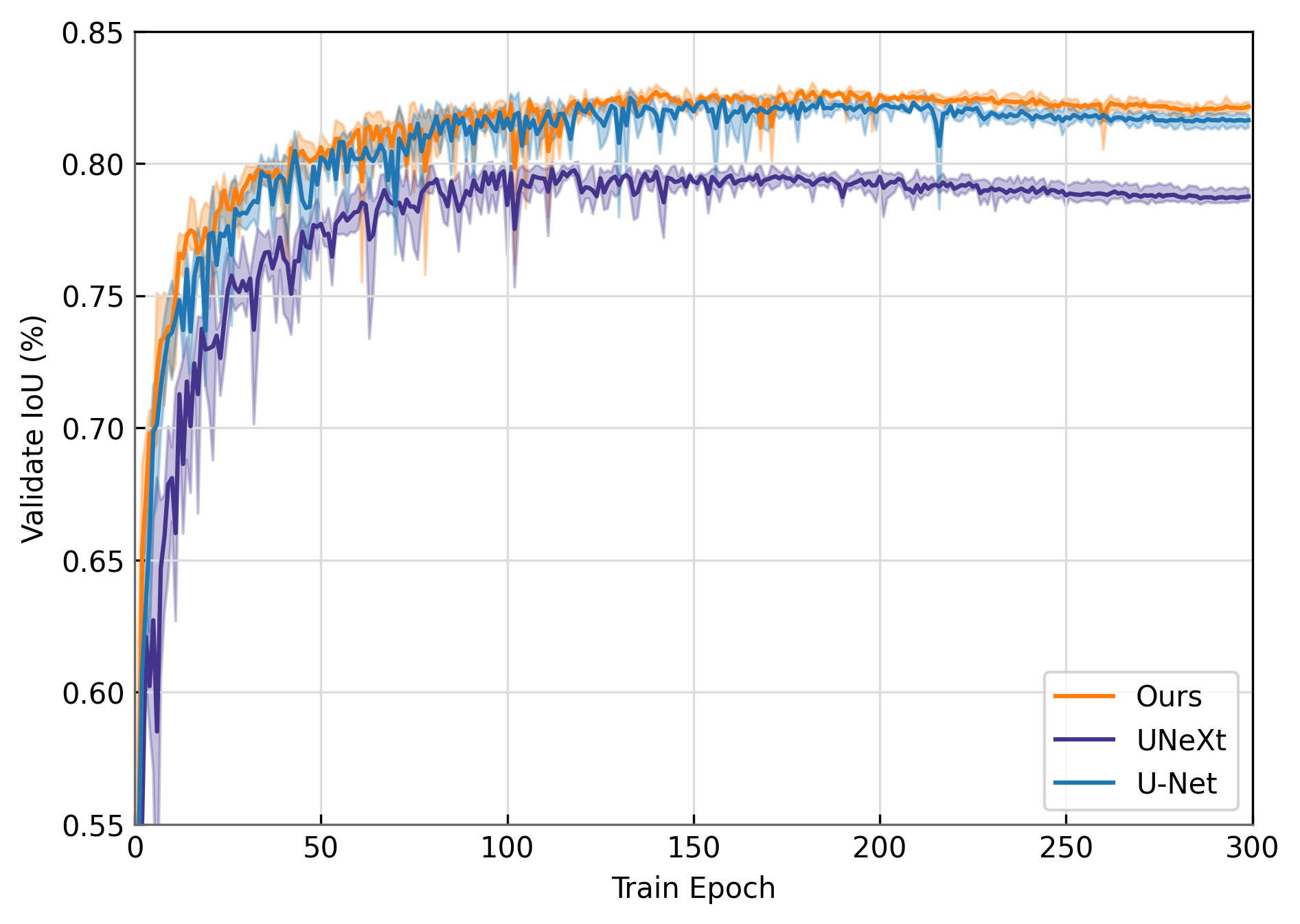}}
  \centerline{\scriptsize{(c) TUS}}\medskip
\end{minipage}
\begin{minipage}[b]{0.495\linewidth}
  \centering
  \centerline{\includegraphics[width=\columnwidth]{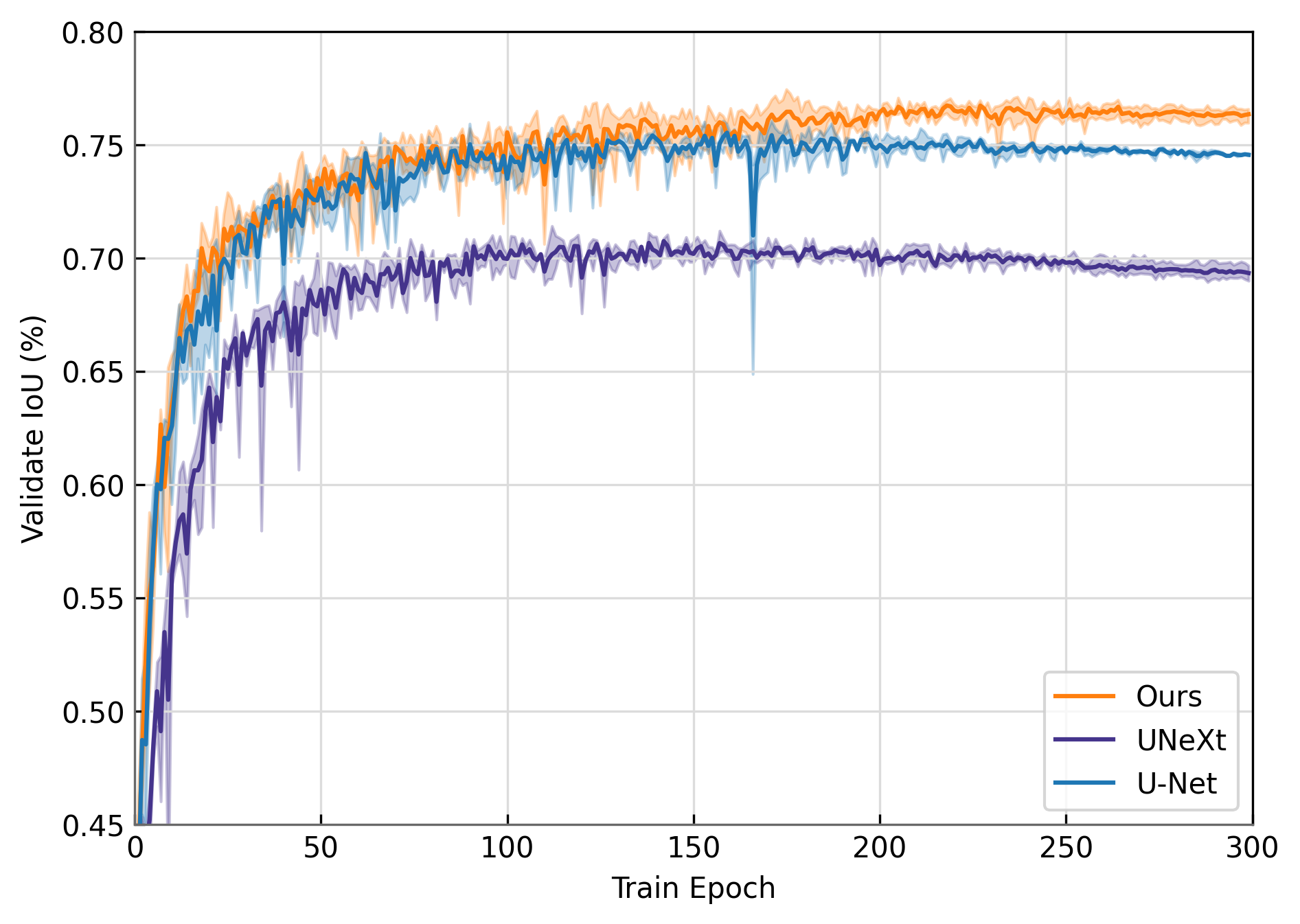}}
  \centerline{\scriptsize{(d) TUSCUI}}\medskip
\end{minipage}

\caption{Comparison Training Stability.}
\label{fig4}
\vspace{-0.5cm}
\end{figure}

\section{Conclusion}
In this work, we propose a novel lightweight medical image segmentation network, CMUNeXt, that achieves an optimal trade-off between segmentation performance and computation consumption through a refined network design. Our extensive experiments demonstrate that the inductive bias and large kernel of convolutions in light-weight networks is crucial in fitting scarce medical data and achieving high segmentation performance in medical image. 

Furthermore, we offer three different variants of the CMUNeXt architecture, each tailored to meet the environment and performance requirements for deployment on different edge and device platforms. These variants enable fast and accurate auxiliary diagnosis in various real-world scenarios.

Overall, our proposed CMUNeXt architecture provides a promising solution for medical image segmentation tasks, with the potential to significantly improve diagnostic accuracy and efficiency.

\appendix

\section{Acknowledgments}
This work is supported by Shandong Natural Science Foundation of China (ZR2020MH290) and by the Joint Funds of the National Natural Science Foundation of China (U22A2033).

\bibliography{aaai24}

\begin{thebibliography}{23}
\providecommand{\natexlab}[1]{#1}

\bibitem[{Al-Dhabyani et~al.(2020)Al-Dhabyani, Gomaa, Khaled, and Fahmy}]{busi}
Al-Dhabyani, W.; Gomaa, M.; Khaled, H.; and Fahmy, A. 2020.
\newblock Dataset of breast ultrasound images.
\newblock \emph{Data in brief}, 28: 104863.

\bibitem[{Cao et~al.(2022)Cao, Wang, Chen, Jiang, Zhang, Tian, and
  Wang}]{swinunet}
Cao, H.; Wang, Y.; Chen, J.; Jiang, D.; Zhang, X.; Tian, Q.; and Wang, M. 2022.
\newblock Swin-unet: Unet-like pure transformer for medical image segmentation.
\newblock In \emph{European conference on computer vision}, 205--218. Springer.

\bibitem[{Chen et~al.(2021)Chen, Lu, Yu, Luo, Adeli, Wang, Lu, Yuille, and
  Zhou}]{transunet}
Chen, J.; Lu, Y.; Yu, Q.; Luo, X.; Adeli, E.; Wang, Y.; Lu, L.; Yuille, A.~L.;
  and Zhou, Y. 2021.
\newblock Transunet: Transformers make strong encoders for medical image
  segmentation.
\newblock \emph{arXiv preprint arXiv:2102.04306}.

\bibitem[{Ding et~al.(2022)Ding, Zhang, Han, and Ding}]{31x31}
Ding, X.; Zhang, X.; Han, J.; and Ding, G. 2022.
\newblock Scaling up your kernels to 31x31: Revisiting large kernel design in
  cnns.
\newblock In \emph{Proceedings of the IEEE/CVF conference on computer vision
  and pattern recognition}, 11963--11975.

\bibitem[{Dosovitskiy et~al.(2020)Dosovitskiy, Beyer, Kolesnikov, Weissenborn,
  Zhai, Unterthiner, Dehghani, Minderer, Heigold, Gelly et~al.}]{vit}
Dosovitskiy, A.; Beyer, L.; Kolesnikov, A.; Weissenborn, D.; Zhai, X.;
  Unterthiner, T.; Dehghani, M.; Minderer, M.; Heigold, G.; Gelly, S.; et~al.
  2020.
\newblock An image is worth 16x16 words: Transformers for image recognition at
  scale.
\newblock \emph{arXiv preprint arXiv:2010.11929}.

\bibitem[{Han, Jian, and Wang(2022)}]{convunext}
Han, Z.; Jian, M.; and Wang, G.-G. 2022.
\newblock ConvUNeXt: An efficient convolution neural network for medical image
  segmentation.
\newblock \emph{Knowledge-Based Systems}, 253: 109512.

\bibitem[{He et~al.(2016)He, Zhang, Ren, and Sun}]{resnet}
He, K.; Zhang, X.; Ren, S.; and Sun, J. 2016.
\newblock Deep residual learning for image recognition.
\newblock In \emph{Proceedings of the IEEE conference on computer vision and
  pattern recognition}, 770--778.

\bibitem[{Huang et~al.(2020)Huang, Lin, Tong, Hu, Zhang, Iwamoto, Han, Chen,
  and Wu}]{unet3+}
Huang, H.; Lin, L.; Tong, R.; Hu, H.; Zhang, Q.; Iwamoto, Y.; Han, X.; Chen,
  Y.-W.; and Wu, J. 2020.
\newblock Unet 3+: A full-scale connected unet for medical image segmentation.
\newblock In \emph{ICASSP 2020-2020 IEEE international conference on acoustics,
  speech and signal processing (ICASSP)}, 1055--1059. IEEE.

\bibitem[{Liu et~al.(2021)Liu, Lin, Cao, Hu, Wei, Zhang, Lin, and Guo}]{swin}
Liu, Z.; Lin, Y.; Cao, Y.; Hu, H.; Wei, Y.; Zhang, Z.; Lin, S.; and Guo, B.
  2021.
\newblock Swin transformer: Hierarchical vision transformer using shifted
  windows.
\newblock In \emph{Proceedings of the IEEE/CVF international conference on
  computer vision}, 10012--10022.

\bibitem[{Liu et~al.(2022)Liu, Mao, Wu, Feichtenhofer, Darrell, and
  Xie}]{convnext}
Liu, Z.; Mao, H.; Wu, C.-Y.; Feichtenhofer, C.; Darrell, T.; and Xie, S. 2022.
\newblock A convnet for the 2020s.
\newblock In \emph{Proceedings of the IEEE/CVF conference on computer vision
  and pattern recognition}, 11976--11986.

\bibitem[{Mehta and Rastegari(2021)}]{mobilevit}
Mehta, S.; and Rastegari, M. 2021.
\newblock Mobilevit: light-weight, general-purpose, and mobile-friendly vision
  transformer.
\newblock \emph{arXiv preprint arXiv:2110.02178}.

\bibitem[{Oktay et~al.(2018)Oktay, Schlemper, Folgoc, Lee, Heinrich, Misawa,
  Mori, McDonagh, Hammerla, Kainz et~al.}]{attunet}
Oktay, O.; Schlemper, J.; Folgoc, L.~L.; Lee, M.; Heinrich, M.; Misawa, K.;
  Mori, K.; McDonagh, S.; Hammerla, N.~Y.; Kainz, B.; et~al. 2018.
\newblock Attention u-net: Learning where to look for the pancreas.
\newblock \emph{arXiv preprint arXiv:1804.03999}.

\bibitem[{Ronneberger, Fischer, and Brox(2015)}]{unet}
Ronneberger, O.; Fischer, P.; and Brox, T. 2015.
\newblock U-net: Convolutional networks for biomedical image segmentation.
\newblock In \emph{Medical Image Computing and Computer-Assisted
  Intervention--MICCAI 2015: 18th International Conference, Munich, Germany,
  October 5-9, 2015, Proceedings, Part III 18}, 234--241. Springer.

\bibitem[{Sandler et~al.(2018)Sandler, Howard, Zhu, Zhmoginov, and
  Chen}]{mobilenetv2}
Sandler, M.; Howard, A.; Zhu, M.; Zhmoginov, A.; and Chen, L.-C. 2018.
\newblock Mobilenetv2: Inverted residuals and linear bottlenecks.
\newblock In \emph{Proceedings of the IEEE conference on computer vision and
  pattern recognition}, 4510--4520.

\bibitem[{Tang et~al.(2022)Tang, Wang, Ning, Xian, and Ding}]{cmunet}
Tang, F.; Wang, L.; Ning, C.; Xian, M.; and Ding, J. 2022.
\newblock CMU-Net: A Strong ConvMixer-based Medical Ultrasound Image
  Segmentation Network.
\newblock \emph{arXiv preprint arXiv:2210.13012}.

\bibitem[{Tolstikhin et~al.(2021)Tolstikhin, Houlsby, Kolesnikov, Beyer, Zhai,
  Unterthiner, Yung, Steiner, Keysers, Uszkoreit et~al.}]{mlpmixer}
Tolstikhin, I.~O.; Houlsby, N.; Kolesnikov, A.; Beyer, L.; Zhai, X.;
  Unterthiner, T.; Yung, J.; Steiner, A.; Keysers, D.; Uszkoreit, J.; et~al.
  2021.
\newblock Mlp-mixer: An all-mlp architecture for vision.
\newblock \emph{Advances in neural information processing systems}, 34:
  24261--24272.

\bibitem[{Trockman and Kolter(2022)}]{convmixer}
Trockman, A.; and Kolter, J.~Z. 2022.
\newblock Patches are all you need?
\newblock \emph{arXiv preprint arXiv:2201.09792}.

\bibitem[{Valanarasu et~al.(2021)Valanarasu, Oza, Hacihaliloglu, and
  Patel}]{medT}
Valanarasu, J. M.~J.; Oza, P.; Hacihaliloglu, I.; and Patel, V.~M. 2021.
\newblock Medical transformer: Gated axial-attention for medical image
  segmentation.
\newblock In \emph{Medical Image Computing and Computer Assisted
  Intervention--MICCAI 2021: 24th International Conference, Strasbourg, France,
  September 27--October 1, 2021, Proceedings, Part I 24}, 36--46. Springer.

\bibitem[{Valanarasu and Patel(2022)}]{unext}
Valanarasu, J. M.~J.; and Patel, V.~M. 2022.
\newblock Unext: Mlp-based rapid medical image segmentation network.
\newblock In \emph{International Conference on Medical Image Computing and
  Computer-Assisted Intervention}, 23--33. Springer.

\bibitem[{Vaswani et~al.(2017)Vaswani, Shazeer, Parmar, Uszkoreit, Jones,
  Gomez, Kaiser, and Polosukhin}]{attention}
Vaswani, A.; Shazeer, N.; Parmar, N.; Uszkoreit, J.; Jones, L.; Gomez, A.~N.;
  Kaiser, {\L}.; and Polosukhin, I. 2017.
\newblock Attention is all you need.
\newblock \emph{Advances in neural information processing systems}, 30.

\bibitem[{Xue et~al.(2021)Xue, Zhu, Fu, Hu, Li, Zhang, and Heng}]{global}
Xue, C.; Zhu, L.; Fu, H.; Hu, X.; Li, X.; Zhang, H.; and Heng, P.-A. 2021.
\newblock Global guidance network for breast lesion segmentation in ultrasound
  images.
\newblock \emph{Medical image analysis}, 70: 101989.

\bibitem[{Zhang et~al.(2022)Zhang, Xian, Cheng, Shareef, Ding, Xu, Huang,
  Zhang, Ning, and Wang}]{busis}
Zhang, Y.; Xian, M.; Cheng, H.-D.; Shareef, B.; Ding, J.; Xu, F.; Huang, K.;
  Zhang, B.; Ning, C.; and Wang, Y. 2022.
\newblock BUSIS: a benchmark for breast ultrasound image segmentation.
\newblock In \emph{Healthcare}, volume~10, 729. MDPI.

\bibitem[{Zhou et~al.(2019)Zhou, Siddiquee, Tajbakhsh, and Liang}]{unet++}
Zhou, Z.; Siddiquee, M. M.~R.; Tajbakhsh, N.; and Liang, J. 2019.
\newblock Unet++: Redesigning skip connections to exploit multiscale features
  in image segmentation.
\newblock \emph{IEEE transactions on medical imaging}, 39(6): 1856--1867.

\end{thebibliography}

\end{document}